\documentclass[runningheads]{llncs}
\usepackage{graphicx}
\usepackage{algorithm}
\usepackage{comment}
\usepackage[noend]{algpseudocode}
\usepackage{xcolor}
\newcommand{\etal}{\textit{et al.}}

\begin{document}
\title{Study Group Learning: Improving Retinal Vessel Segmentation Trained with Noisy Labels }
\titlerunning{SGL for Retinal Vessel Segmentation}
%
\author{Yuqian Zhou\inst{1} \and
Hanchao Yu\inst{1}\and
Humphrey Shi\inst{1,2}}
%
\authorrunning{Zhou et al.}
%
\institute{University of Illinois at Urbana-Champaign \and
University of Oregon\\
}

%

\maketitle              
\begin{abstract}
Retinal vessel segmentation from retinal images is an essential task for developing the computer-aided diagnosis system for retinal diseases. Efforts have been made on high-performance deep learning-based approaches to segment the retinal images in an end-to-end manner. However, the acquisition of retinal vessel images and segmentation labels requires onerous work from professional clinicians, which results in smaller training dataset with incomplete labels. As known, data-driven methods suffer from data insufficiency, and the models will easily over-fit the small-scale training data. Such a situation becomes more severe when the training vessel labels are incomplete or incorrect. In this paper, we propose a Study Group Learning (SGL) scheme to improve the robustness of the model trained on noisy labels. Besides, a learned enhancement map provides better visualization than conventional methods as an auxiliary tool for clinicians. Experiments demonstrate that the proposed method further improves the vessel segmentation performance in DRIVE and CHASE$\_$DB1 datasets, especially when the training labels are noisy. Our code is available at {\color{magenta} \url{https://github.com/SHI-Labs/SGL-Retinal-Vessel-Segmentation}}.

\end{abstract}
\section{Introduction}
Retinal inspection is an effective approach for the diagnose of multiple retinal diseases including diabetic retinopathy, epiretinal membrane, retinal detachment, retinal tear \textit{etc}. Among them, retinal vacular disorders which affects retinal blood vessels are usually caused by other medical diseases like artherosclerosis, hypertension, or human blood circulation problems \cite{brand2012management}. Those disorders will severely influence human's vision functions and cause obvious symptoms, but can be effectively diagnosed and analyzed by retinal vessel inspection in the collected fundus images. Advanced medical imaging system makes it possible to obtain high-resolution fundus images. However, in practical medical services, visual inspection may still require the involvement and tedious work of neurologists, cardiologists, ophthalmologists, and other experts in retinal vascular diseases. To release their burden on screening multiple diseased retino from thousands even millions of healthy retinos, an automatic and high-performance Computer-Aided Diagnosis (CAD) system is desirable to conduct pre-screening and other auxiliary works. Specifically for retinal vascular diseases, we expect the system to provide high-quality enhanced images for a better visualization, and reasonable segmentation of the vessel patterns from the complex and noisy images.   

Plenty of previous efforts have been made in automatic retinal vessel segmentation. Conventionally, hand-crafted filters \cite{niemeijer2004comparative,oloumi2007detection,ricci2007retinal,you2011segmentation} like Gabor \cite{oloumi2007detection} and Gaussian-based ones \cite{niemeijer2004comparative} are explored to extract features for pixel selection, vessel clustering and segmentation. Recently, data-driven based methods utilize UNet-based model \cite{ronneberger2015u} or its variants \cite{wang2020rvseg,xu2020boosting,zhang2020befd,zhang2020retinal,lan2020elastic} to achieve significant performance compared with traditional methods. Those deep learning methods focus on the design of UNet structures with better feature representation \cite{zhang2020befd,wang2020rvseg}, or the decouple of structure and textures of retinal images \cite{zhang2020retinal}. However, data-driven methods highly suffer from over-fitting issues when the given training data is insufficient. Previously proposed methods cannot overcome the issues of small-scale training data with noisy labels given by the clinicians. 

Effectively training networks with noisy labels \cite{song2020learning} is a rigid need in industry and an interesting task in academia. Previous research works mostly focus on image classification tasks and develop methodologies like optimizing robust loss functions \cite{ghosh2017robust,zhang2018generalized}, regularizing labels \cite{goodfellow2014explaining,pereyra2017regularizing}, or actively selecting samples \cite{jiang2018mentornet,han2018co} \textit{etc.}. However, noisy pixel-level labels existing in segmentation tasks are not well-studied, especially for medical image processing tasks. Shu \textit{et al.} proposed a LVC-Net \cite{shu2019lvc} to adjust the incorrect pixel-wise labels via a deformable spatial transformation module guided by low-level visual cues. But their method cannot be applied to retinal blood vessel images, because the entire small blood vessels may be mislabeled and cannot be corrected via spatial transformation. Xue \textit{et al.} \cite{xue2020cascaded} studied a multi-stage training framework with sample selection for chest X-ray images. They synthesized noisy annotations with image dilation and erosion. However, due to the thickness of blood vessels and less training data compared to other medical image data, neither label synthesis nor sampling procedures are suitable for blood vessel images. Considering the characteristics of blood vessel images, we introduce a novel noisy label synthesis pipeline for retinal vessel images, and propose a Study-Group Learning (SGL) framework to improve the model robustness on noisy labels. 

In this paper, we mainly study the two main practical problems for retinal vessel segmentation task. First, we explore the deeply unsupervised learned enhancement of the original retinal images compared with traditional contrast adjustment methods like CLAHE \cite{pizer1987adaptive}. Second, suppose the ground truth segmentation labels given by the clinicians are incomplete and noisy, which yields missing annotations of some vessel segments, we study the effective learning scheme to improve the robustness of the model while training on noisy labels. Therefore, the contributions can be summarized in the following aspects. 
\begin{itemize}
  \item First, to better visualize the retinal image and understand the model, we design the network to output the learned enhancement map. Compared with other baselines, the high-contrast learned map are better visually plausible, and provides a better auxiliary tool to aid clinicians for visual inspection or manual segmentation.
  \item Second, we design a pipeline to manipulate vessel segmentation labels. Given a complete annotations, the proposed approach simulates to automatically erase some vessel segments. 
  \item Third, we propose a Study Group Learning (SGL) framework to improve the generalization ability of the learned model, and better address the missing annotation problems in the training set. The model achieves a more robust performance even some of the vessel pixels are mislabeled.
\end{itemize}


\section{Methodology}
\subsection{Model Structure}
\begin{figure}[t]
\centering
  \includegraphics[width=1\linewidth]{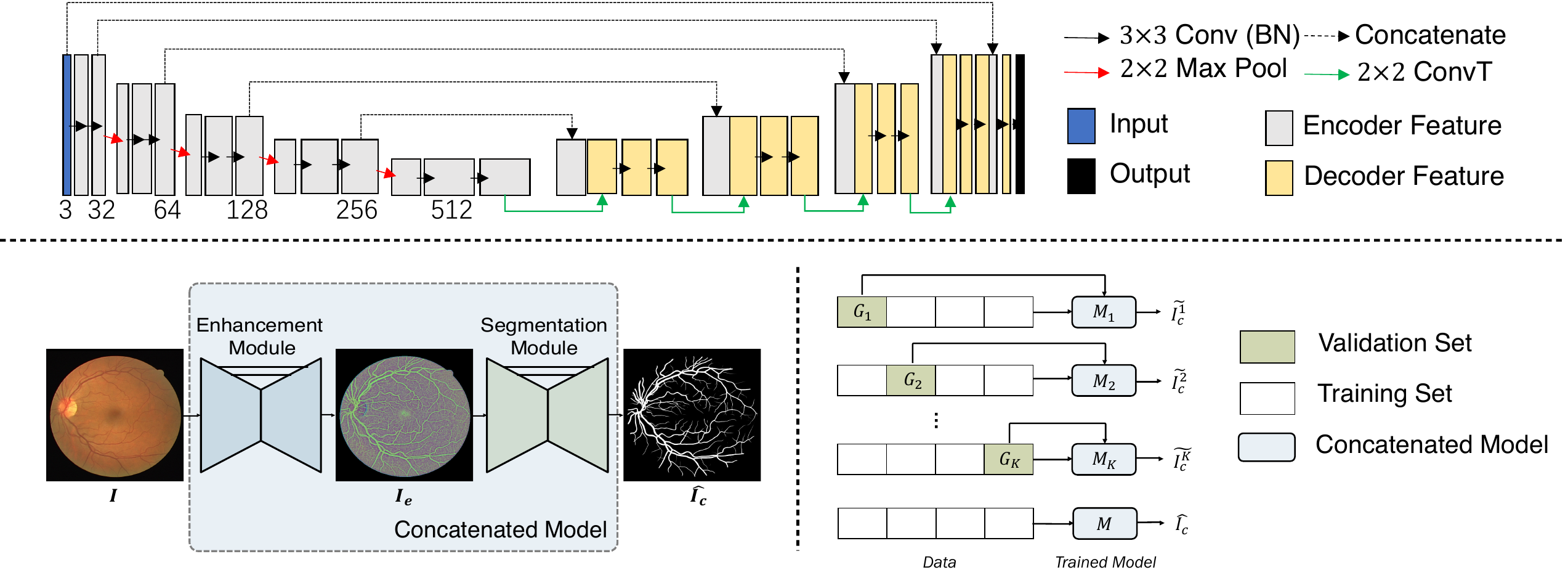}
\caption{Baseline model structure and the proposed Study Group Learning (SGL) scheme. The baseline network is a concatenated UNet consisting of an enhancement module and a segmentation module. The three-channel enhancement map $I_e$ is obtained from the bottleneck structure. The SGL is inspired by the cross-validation training scheme and knowledge distillation idea. We first split the whole training set $G$ into $K$ subsets $\{G_k\}$ and feed the model $M_k$ with $G_k$. The obtained estimation $\tilde{I_c^k}$ of $G_k$ is utilized as the pseudo labels for joint optimization.}
\label{fig:model}
\end{figure}
The proposed baseline model structure is illustrated in Fig. \ref{fig:model}. We utilize the concatenated UNet consisting of the enhancement and segmentation modules to learn both the enhancement and the segmentation map. Different from previous works, we do not pre-process the retinal images, but directly utilize the raw captured images to preserve the entire information. Specifically, given a three-channel raw retinal image $I$ as the input, we aim to process the image by enhancing its contrast and highlighting the vessel structures in $I_e$, and estimate the segmentation map $\hat{I_c}$ of the vessels matched with the ground truth segmentation map $I_c$ given by professional clinicians. Notice that $I_e$ preserves the maximum image contents including the vessel structures and retinal textures. It helps the clinicians to inspect the segmentation results $\hat{I_c}$ and better explains the learned model. 

During testing, we enhance the raw image and infer the segmentation maps from unseen retinal images. 
To cope with noisy labels in vessel segmentation datasets, we propose and follow a Study Group Learning (SGL) scheme to train the baseline model, which is explained in the following subsections. 

\subsection{Study Group Learning (SGL) Scheme}
Tasks like retinal vessel segmentation faces the problem of small-scale dataset and incorrect or incomplete vessel annotations. These two practical problems will make the deeply trained model very easily over-fitting the training set. The severe over-fitting problem does harm to the generalization ability and robustness of the model to unseen testing data. In addition to conventional data augmentation approaches like image transformation and random warping, we propose to alternate the training scheme. Inspired by K-fold cross-validation scheme and knowledge distillation \cite{hinton2015distilling} approaches, we propose the K-fold Study Group Learning (SGL) to better cope with noisy labels in small datasets, especially for the retinal vessel segmentation task.

The pipeline is illustrated in Fig. \ref{fig:model}. Specifically, we first randomly and averagely split the whole training set $(G, I_c)$ into $K$ subsets $\{(G_k, I_{ck})\}$, where $k \in [1, K]$. Like the cross-validation scheme, we train totally $K$ models $\{M_k\}$. For each $M_k$, we train it on $(G\backslash G_k, I_c\backslash I_{ck})$ using pixel-wise binary cross entropy loss. 
After optimizing the model set $\{M_k\}$ as $\{M_k^*\}$, we infer the estimated segmentation label $\tilde{I_{ck}}$ of $G_k$ as the pseudo label, where
 \begin{equation}
     \tilde{I_{ck}} = M_k^* (G_k), k\in [1,K].
 \end{equation}
Finally, we train a model $M$ from scratch by jointly optimizing the ground truth vessel labels $I_c$ and the obtained pseudo label set $\tilde{I_c} = \bigcup_{k=1}^{K}\tilde{I_{ck}}$ as,
 \begin{equation}
     \mathcal{L}_{SGL} = CE(\hat{I_c}, I_c) + \lambda CE(\hat{I_c}, \tilde{I_c}),
\label{eq:loss}
 \end{equation}
where $\hat{I_c} = M(G)$, $CE$ is the cross entropy loss, and $\lambda=1$.
\begin{figure}[t]
\centering
  \includegraphics[width=1\linewidth]{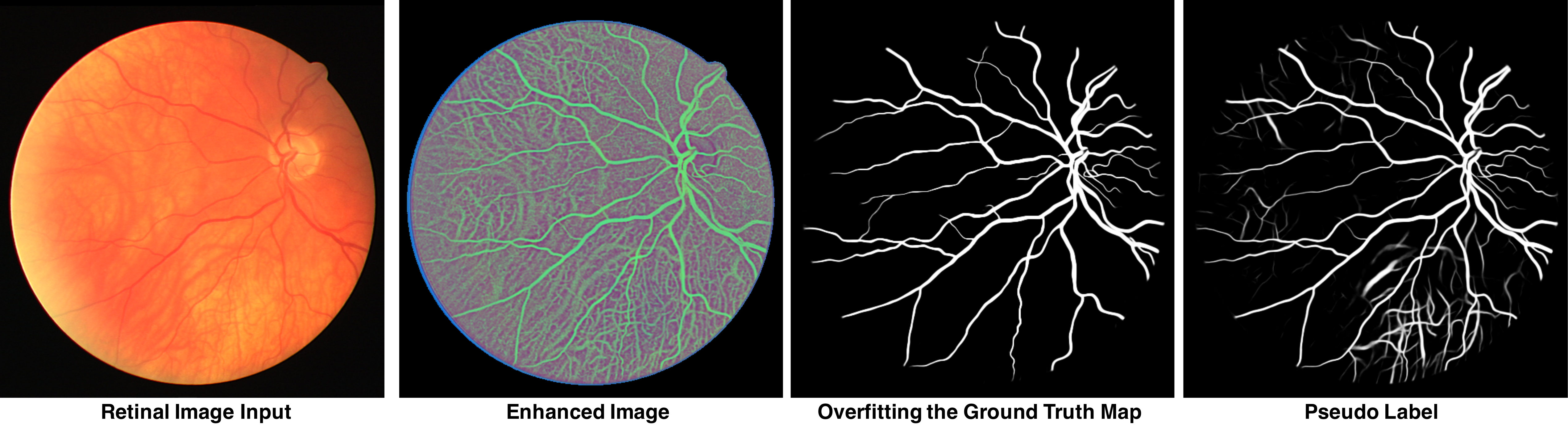}
\caption{Inference results on one training example. Clinicians may only label some salient vessels while ignoring the ambiguous ones. The model trained on the entire set will over-fit the ground truth annotations. However, inference results from the model trained on the subset can serve as the pseudo label for model regularization purpose. }
  \label{fig:inf}
\end{figure}

Intuitively, we name the proposed learning scheme as Study Group Learning (SGL) because each model $M_k$ trained with partial training set can be regarded as a 'Study Group'. The final model $M$ is a process of group discussion by merging and fusing the knowledge from different study groups. Theoretically, the second term in Equ. \ref{eq:loss} can be regarded as a regularization to avoid over-fitting the ground-truth labels, especially when the given labels contain some noises and possibly incorrect annotations. Fig. \ref{fig:inf} shows one example of such cases. In this example, we set $K=2$, and the inference of the training samples $I$ from $M_1$ and $M_2$ are shown in the right two columns, where $I$ is in the training set of $M_1$ but not in the training partition of $M_2$. $M_1$ highly overfits the given labels of $I$ by ignoring multiple thin and ambiguous vessels. However, if not trained on $I$, $M_2$ can infer some of the ignored vessels as the pseudo labels. While combining these two labels, the final model $M$ can intuitively learn better representation and become more generalized. 

\subsection{Vessel Label Erasing}
\begin{figure}[t]
\centering
  \includegraphics[width=1\linewidth]{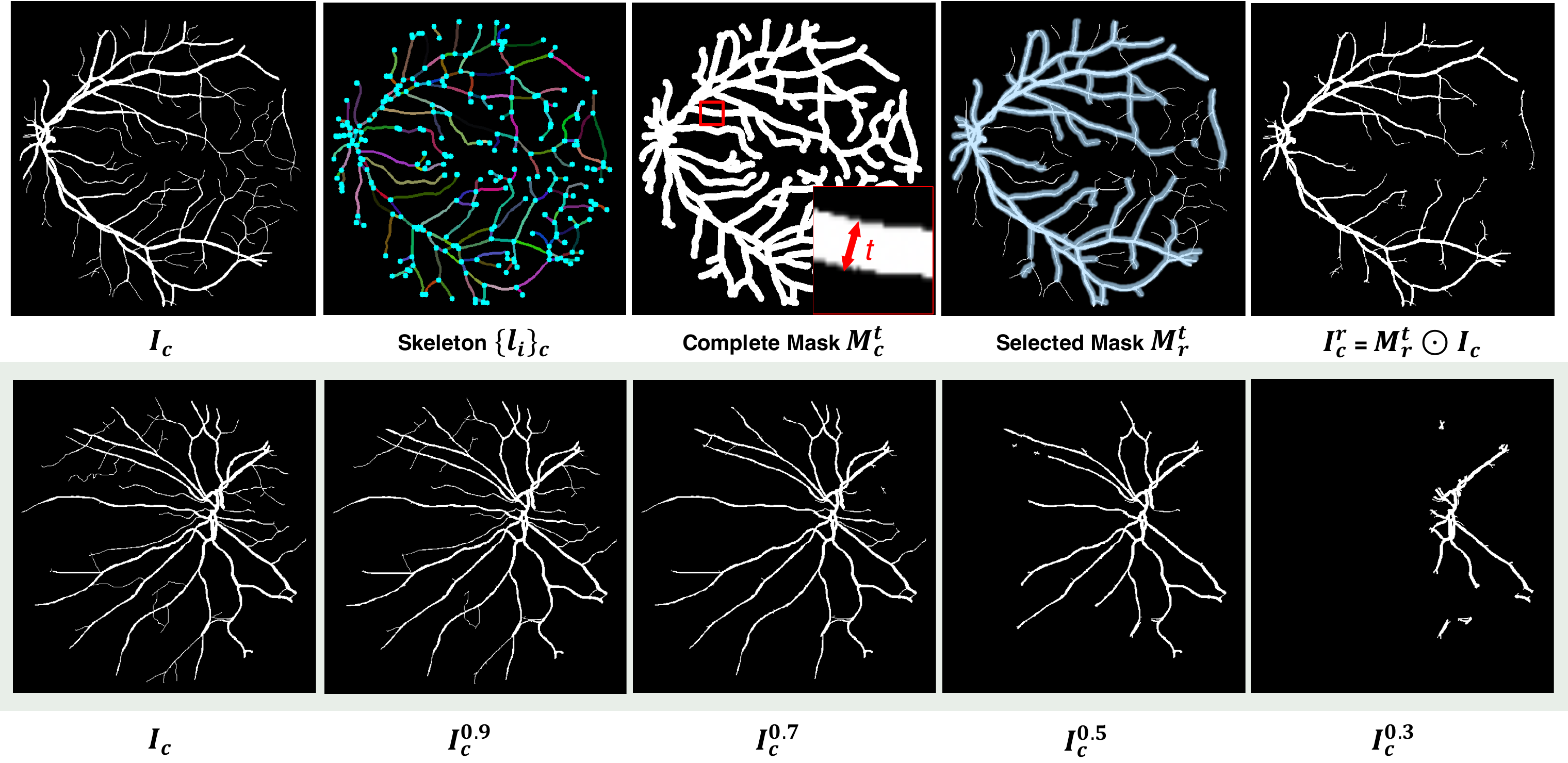}
\caption{Vessel label erasing process. Given the complete label map $I_c$, we compute the skeleton of the vessels, and dilate the vessel skeleton to mask by drawing the polylines with width $t$. We then rank the vessel segments by their approximated thickness, and include the thick vessels according to ratio $r$. The second row shows one instance of label erasing with $r$ from 1 to 0.3.}
\label{fig:synp}
\end{figure}
Annotating the retinal vessels requires the involvement of professional clinicians, and the process of manual labeling is onerous, which reveals one of the reasons why public retinal vessel databases are always small-scale or partially-labeled. It is also common that some labels of thinner vessels are missing due to the annotators' errors. To resemble this practical situation in industry, we propose to synthesize an incomplete map ${I_c^r}$ by erasing some labeled vessel segments $I_r$ from the ground truth segmentation map $I_c = I_c^r \cup I_r$, where $r$ is the removal ratio. The process is illustrated in Fig. \ref{fig:synp}.

To generate $I_c^r$, we first compute and approximate the skeleton of $I_c$ using the method proposed by Zhang \etal \cite{zhang1984fast} followed by a skeleton tracing approach \footnote{skeleton-tracing: https://github.com/LingDong-/skeleton-tracing}. This algorithm converts the binary segmentation map into a set of polylines $L_c = \{l_i\}_c$, where $i \in [1, M]$ and each $l_i$ is stored as an array of coordinates \textit{i.e.} ${l_i = \{p_d\}}$. The geometric polylines and their spatial relationship represent the topological skeleton of the annotated vessels. We utilize it to locate the vessel center lines, and roughly compute the thickness of the vessels. 

Second, to compute and rank the thickness of each vessel segment, we redraw the polylines with larger width $t$ on a black canvas to form the complete mask $M_c^t$. Notice that $t$ is the least value which makes $M_c^t$ cover the entire vessel regions in $I_c$. For each $l_i$, the corresponding pixel regions covered by the drawing lines of width $t$ is denoted by $l_i^t$, then the thickness $s_i$ of the vessel segment is measured by $s_i = \frac{\|I_r \cap l_i^t\|}{\|l_i^t\|}$. We rank the polylines according to their thickness $s_i$, and include the thickest vessels in order in the partial set $L_c^r = \{l_j\}$, where $j \in [1, N]$ and $N\leq M$. In $L_c^r$, we include the top-$r$ thick vessels. Finally, we form the selected mask $M_r^t$ from $L_c^r$, and the synthetic segmentation mask ablating some of the thin vessels with ratio $r$ can be computed by $I_c^r = M_r^t \odot I_c$. Fig. \ref{fig:synp} shows one example of $I_c^r$ where r is from 1 to 0.3. Adding a small portions of thin vessels helps the model maintain the ability of segmenting smaller objects. Therefore, we also randomly select $50 \%$ of thin vessels in the set and add them to $L_c^r$. 
\section{Experiments}
\subsection{Dataset and Implementation}
\textbf{DRIVE:} The Digital Retinal Images for Vessel Extraction (DRIVE) \cite{staal2004ridge} dataset consists of 40 images of size $565 \times 584$. The training and testing set are fixed, and the ground truth manual annotations are given. We do not resize the image to alternate the resolution or change the Aspect-Ratio.\\
\textbf{CHASE$\_$DB1:} \cite{fraz2012ensemble} It consists of 28 retinal images of size $999 \times 960$. For a fair comparison with previous methods, we use the first 20 images for training, and the remaining 8 images for testing.

While training the model, we randomly crop the images into $256 \times 256$ patches, and apply data augmentation including horizontal and vertical flip, rotation, transpose, and random elastic warping \cite{simard2003best}. While testing, suppose the image size is $W \times H$, we zero-pad the input image to size $(2^4\times m, 2^4\times m)$, where $2^4 \times m > \max(W, H) > 2^4 \times {(m-1)} $, and crop the estimated map accordingly. Specifically, $m$ is 37 for DRIVE and 63 for CHASE$\_$DB1 dataset. It makes the arbitrary-sized inputs adaptive to the UNet structure with four down-sampling layers, and retain the original resolution and aspect-ratio of the image. 

\subsection{Learned Retinal Image Enhancement}
\begin{figure}[t]
\centering
  \includegraphics[width=1\linewidth]{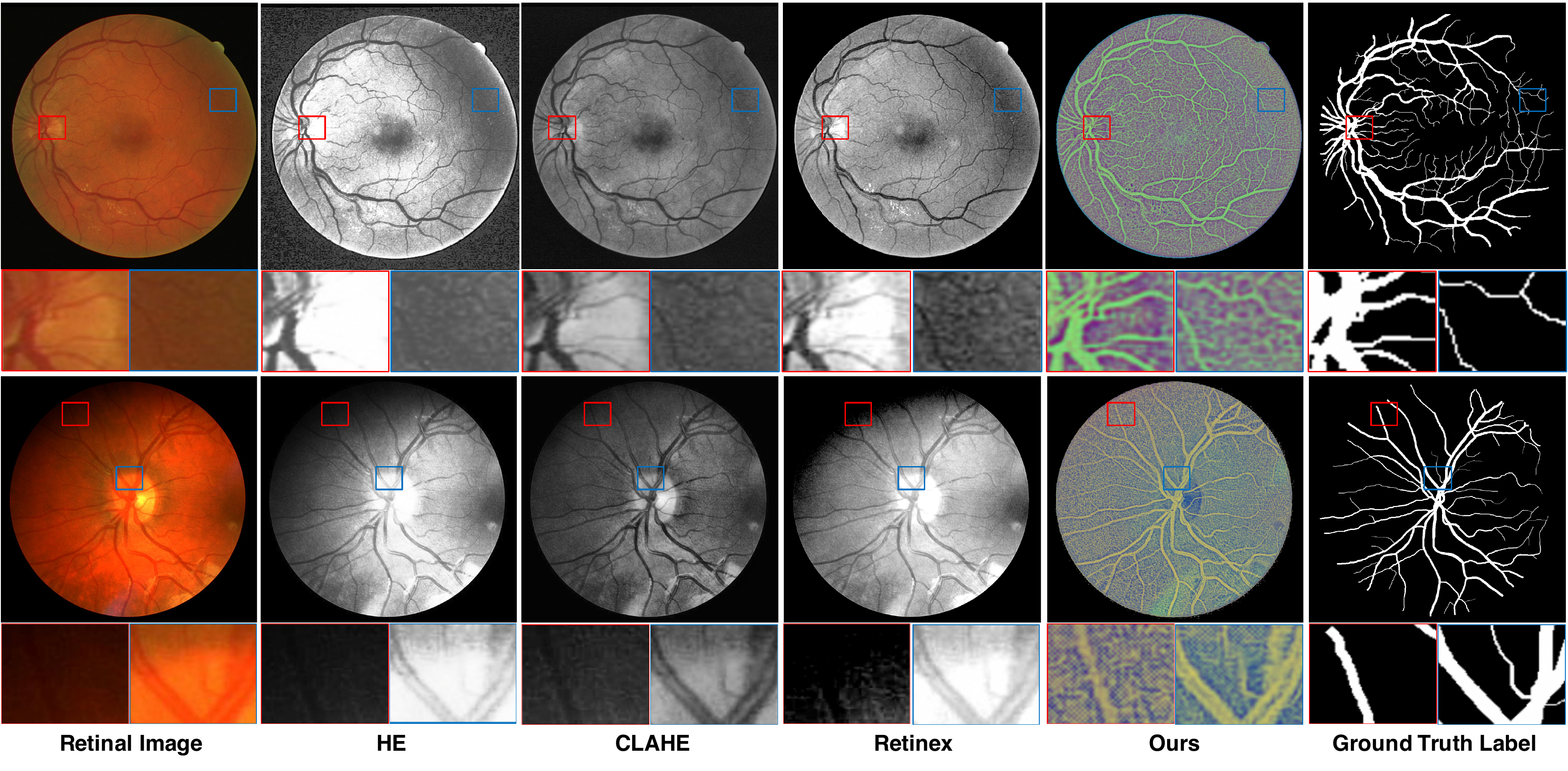}
\caption{The learned enhancement map $I_e$ compared with other baseline methods including Histogram Equalization (HE), Contrast Limited Adaptive Histogram Equalization (CLAHE) 
\cite{pizer1987adaptive}, and Single-scale Retinex \cite{zhao2015retinal}. The learned map demonstrates a better contrast and intensity, enhancing the vessel information for a better identifiable visualization for clinicians. Top row is from DRIVE dataset and the bottom row is from CHASE$\_$DB1 dataset. Zooming in for better visualization. }
  \label{fig:enh}
\end{figure}

\begin{figure}[t]
\setlength{\abovecaptionskip}{0pt}
\centering
  \includegraphics[width=1\linewidth]{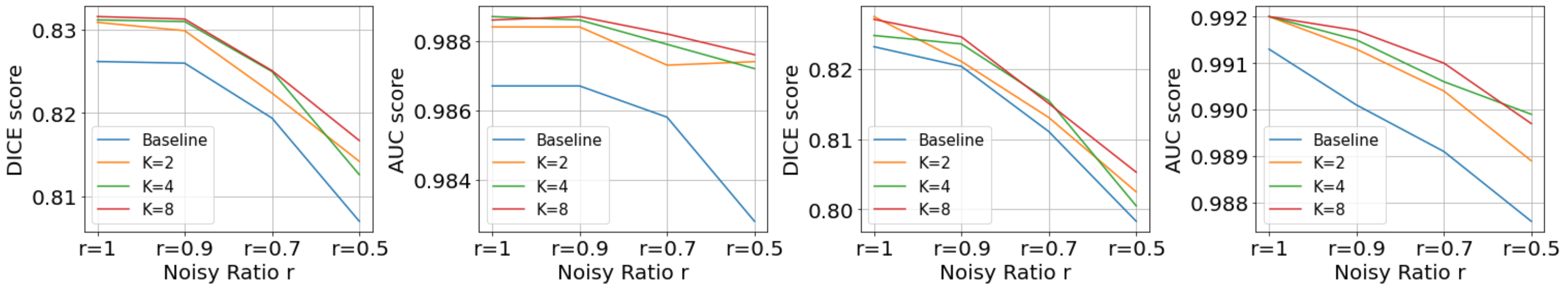}
\caption{The performance of the model in the simulated training set with label noise ratio $r=[1, 0.9, 0.7, 0.5]$. The proposed SGL learning scheme overall improves the robustness in all $K$. Left two columns: DICE and AUC scores on DRIVE. Right two columns: on CHASE$\_$DB1 dataset.}
\label{fig:drive}
\end{figure}
We learn the enhanced map of the original input images by supervising the model on the manual segmentation labels. We extract the bottleneck 3-channel output which can be visualized in a better contrast level. We compare the learned enhancement of the retinal images with other baseline methods including Histogram Equalization (HE), Contrast Limited Adaptive Histogram Equalization (CLAHE) \cite{pizer1987adaptive}, and Retinex \cite{zhao2015retinal}. 

Fig. \ref{fig:enh} shows one example of the visual comparison. Traditional methods like HE, CLAHE and Retinex cannot achieve a uniform contrast level both locally and globally. The cropped patches are from either a brighter or darker regions of the image, making the vessel pixels hard to parse accurately by the inspector. However, the learned map in the fifth column demonstrates a better contrast and intensity level, enhancing the vessel information for a better identifiable visualization for clinicians, especially for the dark regions in CHASE$\_$DB1 images. It highlights the vessel regions while preserving the textures. The obtained enhancement images can also be utilized for visual inspection or labeling.
\subsection{Study Group Learning}

\begin{table}[t]\setlength{\tabcolsep}{5pt}
\centering
\footnotesize
\caption{Comparison with other baseline methods on DRIVE dataset. }
\begin{tabular}{|r|c|c|c|c|c|c|}
\hline
Method & Year & Sensitivity & Specificity & DICE & Accuracy & AUC \\ \hline
R2U-Net \cite{alom2019recurrent}    & 2018 &0.7792 &0.9813 &0.8171 &0.9556 &0.9784\\
LadderNet \cite{zhuang2018laddernet} & 2018 &0.7856 &0.9810 &0.8202 &0.9561 &0.9793\\
Dual E-UNet \cite{wang2019dual}   &2019 &0.7940 &0.9816 &0.8270 &0.9567 &0.9772\\
IterNet \cite{li2020iternet}     &2020 &0.7791 &0.9831 &0.8218 &0.9574 &0.9813\\
SA-UNet \cite{guo2020sa}     &2020 &0.8212 &0.9840 &0.8263 &0.9698 &0.9864\\
BEFD-UNet \cite{zhang2020befd}& 2020 &0.8215 &\bf{0.9845} &0.8267& 0.9701& 0.9867 \\\hline
Our Baseline &2021 &0.8341 &0.9827 &0.8262 &0.9695 &0.9867 \\ 
Our SGL (K=8) &2021&\bf{0.8380} &0.9834 &\bf{0.8316} &\bf{0.9705} &\bf{0.9886} \\\hline

\end{tabular}
\label{exp:drive}
\end{table}

\begin{table}[t]\setlength{\tabcolsep}{5pt}
\centering
\footnotesize
\caption{Comparison with other baseline methods on CHASE$\_$DB1 dataset. }
\begin{tabular}{|r|c|c|c|c|c|c|}
\hline
Method & Year & Sensitivity & Specificity & DICE & Accuracy & AUC \\\hline
R2U-Net \cite{alom2019recurrent}    & 2018 &0.7756 &0.9712 &0.7928&0.9634 &0.9815\\
LadderNet \cite{zhuang2018laddernet} & 2018 &0.7978 &0.9818 &0.8031 &0.9656 &0.9839\\

Dual E-UNet \cite{wang2019dual}   &2019 &0.8074 &0.9821 &0.8037 &0.9661 &0.9812\\
IterNet \cite{li2020iternet}     &2020 &0.7969 &0.9881 &0.8072 &0.9760 &0.9899\\
SA-UNet \cite{guo2020sa}     &2020 &0.8573 &0.9835 &0.8153 &0.9755 &0.9905\\

\hline
Our Baseline &2021 &0.8502 &\bf{0.9854} &0.8232 &0.9769 &0.9913 \\ 
 Our SGL (K=8) &2021&\bf{0.8690} &0.9843 &\bf{0.8271} &\bf{0.9771} &\bf{0.9920} \\\hline
\end{tabular}
\label{exp:chase}
\vspace{-4mm}
\end{table}

Table \ref{exp:drive} and \ref{exp:chase} illustrate the effectiveness of the proposed SGL scheme. Compared with previous works, the proposed learning scheme can boost the DICE score \cite{sorensen1948method} and other evaluation metrics by a large margin. Fig. \ref{fig:drive} shows the DICE and Area Under the Receiver Operating Characteristic (ROC) Curve (AUC) of the model in the simulated training set with label noise ratio $r=[1, 0.9, 0.7, 0.5]$, where $r=1$ represents the original training set. As shown in the figure, erasing some vessel labels in the training set will drastically degrade the system performance, while the SGL learning scheme overall improves the robustness on both datasets. Among all the metrics, AUC does not relate to the thresholding method, indicating a better ability of the model segmenting vessel pixels. Besides, a better sensitivity indicates the model is able to extract more thin vessels and boundary pixels. More results can be found in the supplementary material.

\section{Conclusions}
In this paper, we studied the learning-based retinal vessel segmentation model trained with noisy labels. Specifically, we designed the pipeline of synthesizing noisy labels, and proposed a Study Group Learning (SGL) scheme boosting the performance of model trained with imperfect labels. Besides, the learned enhanced images as a side product made the model explainable and helped the clinicians for visual inspection. We still discovered the gap between models trained with different levels of noisy labels, leaving for future work to further improve the model sensitivity.

\section{Acknowledgement}
This project has been funded by the Jump ARCHES endowment through the Health Care Engineering Systems Center. This work also utilizes resources supported by the National Science Foundation’s Major Research Instrumentation program, grant number 1725729, as well as the University of Illinois at Urbana-Champaign.

\section{Appendix: Quantitative and Qualitative Result}
\begin{table}[h]\setlength{\tabcolsep}{6pt}
\setlength{\abovecaptionskip}{0pt}
\centering
\footnotesize
\caption{Results of different label missing ratio $r$ and fold number $K$ on DRIVE. }
\begin{tabular}{|r|c|c|c|c|c|c|c|}
\hline
r& K & Accuracy & AUC & Sensitivity & Specificity &   DICE (F1) &Vessel IoU  \\ \hline\hline

1 &1&0.9695&0.9867&0.8341&0.9827&0.8262&0.7041\\ \hline
&2&0.9701&0.9884&0.8432&0.9825&0.8309&0.7110\\ \hline
&4&0.9705&0.9887&0.8358&0.9836&0.8312&0.7115\\ \hline
&8&0.9705&0.9886&0.8380&0.9834&0.8316&0.7120\\ \hline\hline
0.9&1&0.9697&0.9867&0.8262&0.9837&0.8260&0.7040\\ \hline
&2&0.9706&0.9884&0.8231&0.9850&0.8299&0.7095\\ \hline
&4&0.9706&0.9886&0.8323&0.9840&0.8310&0.7111\\ \hline
&8&0.9703&0.9887&0.8418&0.9828&0.8313&0.7116\\ \hline\hline
0.7&1&0.9691&0.9858&0.8045&0.9852&0.8194&0.6945\\ \hline
&2&0.9696&0.9873&0.8102&0.9851&0.8224&0.6988\\ \hline
&4&0.9702&0.9879&0.8081&0.9860&0.8250&0.7024\\ \hline
&8&0.9701&0.9882&0.8104&0.9857&0.8251&0.7027\\ \hline
\end{tabular}
\label{exp:drive_full}
\end{table}

\begin{table}[th]\setlength{\tabcolsep}{6pt}
\centering
\footnotesize
\caption{Results of different label missing ratio $r$ and fold number $K$ on CHASE$\_$DB1. }
\begin{tabular}{|r|c|c|c|c|c|c|c|}
\hline
r& K & Accuracy & AUC & Sensitivity & Specificity &   DICE (F1) &Vessel IoU  \\ \hline\hline
1&1&0.9769&0.9913&0.8502&0.9854&0.8232&0.6998\\ \hline
&2&0.9774&0.9920&0.8586&0.9855&0.8275&0.7060\\ \hline
&4&0.9769&0.9920&0.8632&0.9846&0.8248&0.7020\\ \hline
&8&0.9771&0.9920&0.8690&0.9843&0.8271&0.7054\\ \hline\hline
0.9&1&0.9765&0.9901&0.8483&0.9851&0.8204&0.6957\\ \hline
&2&0.9761&0.9913&0.8716&0.9831&0.8211&0.6967\\ \hline
&4&0.9771&0.9915&0.8494&0.9856&0.8236&0.7004\\ \hline
&8&0.9768&0.9917&0.8652&0.9843&0.8246&0.7017\\ \hline\hline
0.7&1&0.9759&0.9891&0.8182&0.9864&0.8111&0.6824\\ \hline
&2&0.9761&0.9904&0.8250&0.9863&0.8131&0.6852\\ \hline
&4&0.9761&0.9906&0.8341&0.9856&0.8155&0.6887\\ \hline
&8&0.9757&0.9910&0.8480&0.9843&0.8151&0.6881\\ \hline
\end{tabular}
\label{exp:chase_full}
\end{table}
\begin{figure}[th]
\centering
  \includegraphics[width=0.9\linewidth]{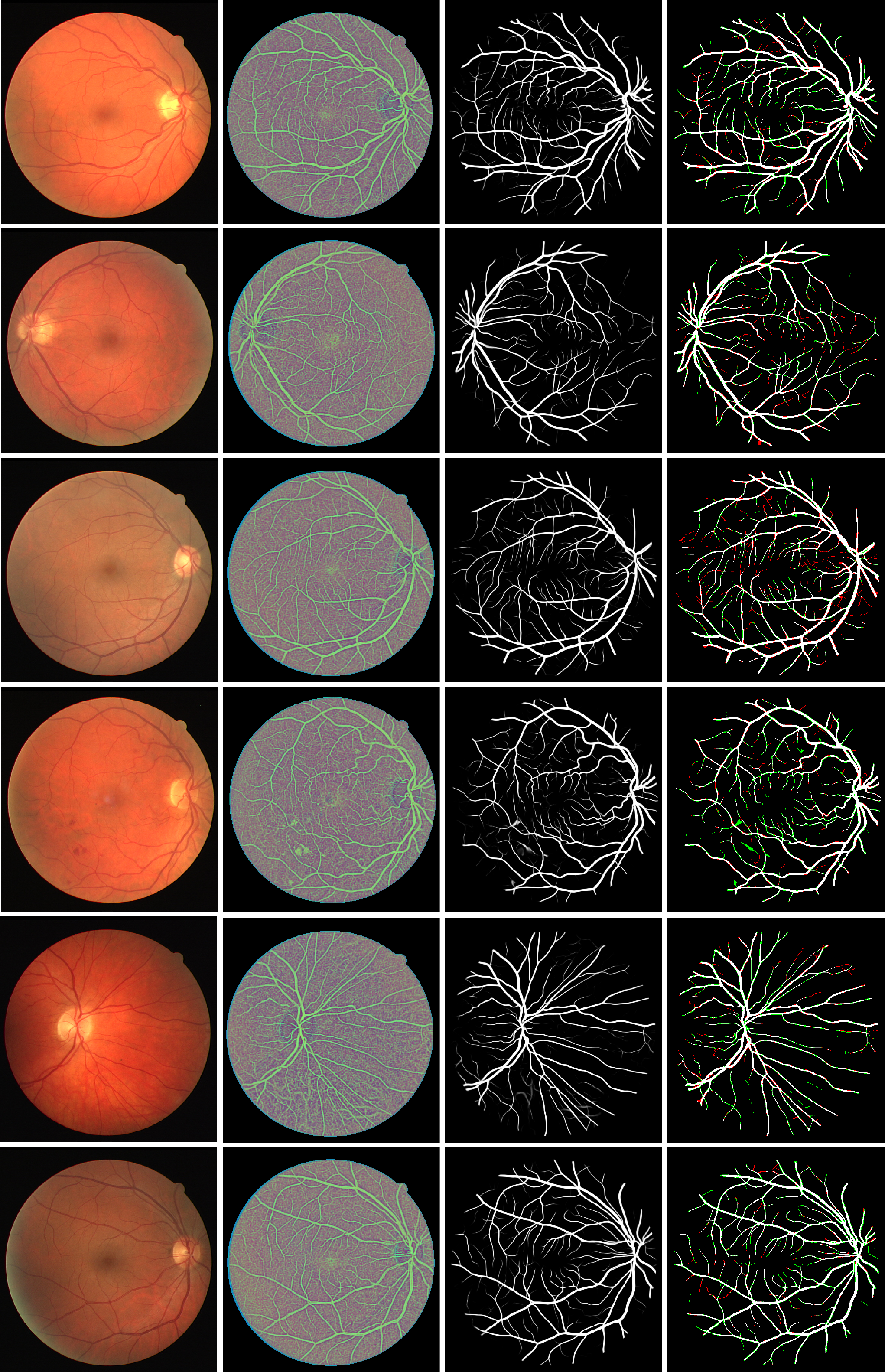}
\caption{Enhancement and segmentation results on DRIVE. From left to right: the raw retinal image, the learned enhanced image, the probability map, and the binary segmentation result. Red color indicates false negative, and green color indicates false positive. Thin vessels are still very challenging to be segmented. }
  \label{fig:drive2}
\end{figure}
\clearpage
\begin{figure}[th]
\centering
  \includegraphics[width=0.9\linewidth]{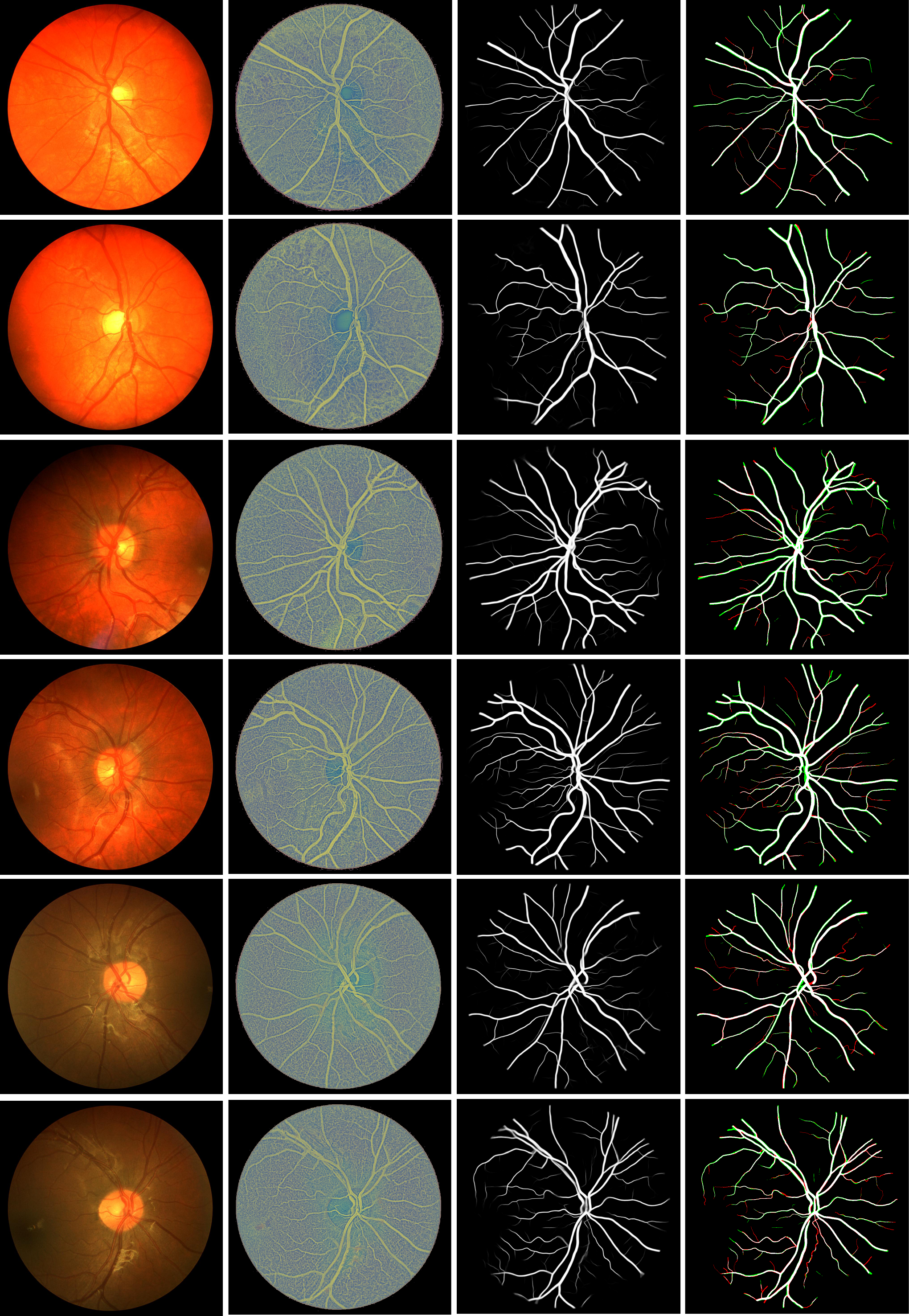}
\caption{Enhancement and segmentation results on CHASE$\_$DB1. From left to right: the raw retinal image, the learned enhanced image, the probability map, and the binary segmentation result. Red color indicates false negative, and green color indicates false positive. }
  \label{fig:chase}
\end{figure}
\clearpage

\bibliographystyle{splncs04}
\bibliography{mybib}

\begin{thebibliography}{10}
\providecommand{\url}[1]{\texttt{#1}}
\providecommand{\urlprefix}{URL }
\providecommand{\doi}[1]{https://doi.org/#1}

\bibitem{alom2019recurrent}
Alom, M.Z., Yakopcic, C., Hasan, M., Taha, T.M., Asari, V.K.: Recurrent
  residual u-net for medical image segmentation. Journal of Medical Imaging
  \textbf{6}(1),  014006 (2019)

\bibitem{brand2012management}
Brand, C.S.: Management of retinal vascular diseases: a patient-centric
  approach. Eye  \textbf{26}(2),  S1--S16 (2012)

\bibitem{fraz2012ensemble}
Fraz, M.M., Remagnino, P., Hoppe, A., Uyyanonvara, B., Rudnicka, A.R., Owen,
  C.G., Barman, S.A.: An ensemble classification-based approach applied to
  retinal blood vessel segmentation. IEEE Transactions on Biomedical
  Engineering  \textbf{59}(9),  2538--2548 (2012)

\bibitem{ghosh2017robust}
Ghosh, A., Kumar, H., Sastry, P.: Robust loss functions under label noise for
  deep neural networks. In: Proceedings of the AAAI Conference on Artificial
  Intelligence. vol.~31 (2017)

\bibitem{goodfellow2014explaining}
Goodfellow, I.J., Shlens, J., Szegedy, C.: Explaining and harnessing
  adversarial examples. arXiv preprint arXiv:1412.6572  (2014)

\bibitem{guo2020sa}
Guo, C., Szemenyei, M., Yi, Y., Wang, W., Chen, B., Fan, C.: Sa-unet: Spatial
  attention u-net for retinal vessel segmentation. arXiv preprint
  arXiv:2004.03696  (2020)

\bibitem{han2018co}
Han, B., Yao, Q., Yu, X., Niu, G., Xu, M., Hu, W., Tsang, I., Sugiyama, M.:
  Co-teaching: Robust training of deep neural networks with extremely noisy
  labels. arXiv preprint arXiv:1804.06872  (2018)

\bibitem{hinton2015distilling}
Hinton, G., Vinyals, O., Dean, J.: Distilling the knowledge in a neural
  network. arXiv preprint arXiv:1503.02531  (2015)

\bibitem{jiang2018mentornet}
Jiang, L., Zhou, Z., Leung, T., Li, L.J., Fei-Fei, L.: Mentornet: Learning
  data-driven curriculum for very deep neural networks on corrupted labels. In:
  International Conference on Machine Learning. pp. 2304--2313. PMLR (2018)

\bibitem{lan2020elastic}
Lan, Y., Xiang, Y., Zhang, L.: An elastic interaction-based loss function for
  medical image segmentation. In: International Conference on Medical Image
  Computing and Computer-Assisted Intervention. pp. 755--764. Springer (2020)

\bibitem{li2020iternet}
Li, L., Verma, M., Nakashima, Y., Nagahara, H., Kawasaki, R.: Iternet: Retinal
  image segmentation utilizing structural redundancy in vessel networks. In:
  The IEEE Winter Conference on Applications of Computer Vision. pp. 3656--3665
  (2020)

\bibitem{niemeijer2004comparative}
Niemeijer, M., Staal, J., van Ginneken, B., Loog, M., Abramoff, M.D.:
  Comparative study of retinal vessel segmentation methods on a new publicly
  available database. In: Medical imaging 2004: image processing. vol.~5370,
  pp. 648--656. International Society for Optics and Photonics (2004)

\bibitem{oloumi2007detection}
Oloumi, F., Rangayyan, R.M., Oloumi, F., Eshghzadeh-Zanjani, P., Ayres, F.J.:
  Detection of blood vessels in fundus images of the retina using gabor
  wavelets. In: 2007 29th Annual International Conference of the IEEE
  Engineering in Medicine and Biology Society. pp. 6451--6454. IEEE (2007)

\bibitem{pereyra2017regularizing}
Pereyra, G., Tucker, G., Chorowski, J., Kaiser, {\L}., Hinton, G.: Regularizing
  neural networks by penalizing confident output distributions. arXiv preprint
  arXiv:1701.06548  (2017)

\bibitem{pizer1987adaptive}
Pizer, S.M., Amburn, E.P., Austin, J.D., Cromartie, R., Geselowitz, A., Greer,
  T., ter Haar~Romeny, B., Zimmerman, J.B., Zuiderveld, K.: Adaptive histogram
  equalization and its variations. Computer vision, graphics, and image
  processing  \textbf{39}(3),  355--368 (1987)

\bibitem{ricci2007retinal}
Ricci, E., Perfetti, R.: Retinal blood vessel segmentation using line operators
  and support vector classification. IEEE transactions on medical imaging
  \textbf{26}(10),  1357--1365 (2007)

\bibitem{ronneberger2015u}
Ronneberger, O., Fischer, P., Brox, T.: U-net: Convolutional networks for
  biomedical image segmentation. In: International Conference on Medical image
  computing and computer-assisted intervention. pp. 234--241. Springer (2015)

\bibitem{shu2019lvc}
Shu, Y., Wu, X., Li, W.: Lvc-net: Medical image segmentation with noisy label
  based on local visual cues. In: International Conference on Medical Image
  Computing and Computer-Assisted Intervention. pp. 558--566. Springer (2019)

\bibitem{simard2003best}
Simard, P.Y., Steinkraus, D., Platt, J.C., et~al.: Best practices for
  convolutional neural networks applied to visual document analysis. In: Icdar.
  vol.~3 (2003)

\bibitem{song2020learning}
Song, H., Kim, M., Park, D., Lee, J.G.: Learning from noisy labels with deep
  neural networks: A survey. arXiv preprint arXiv:2007.08199  (2020)

\bibitem{sorensen1948method}
Sorensen, T.A.: A method of establishing groups of equal amplitude in plant
  sociology based on similarity of species content and its application to
  analyses of the vegetation on danish commons. Biol. Skar.  \textbf{5},  1--34
  (1948)

\bibitem{staal2004ridge}
Staal, J., Abr{\`a}moff, M.D., Niemeijer, M., Viergever, M.A., Van~Ginneken,
  B.: Ridge-based vessel segmentation in color images of the retina. IEEE
  transactions on medical imaging  \textbf{23}(4),  501--509 (2004)

\bibitem{wang2019dual}
Wang, B., Qiu, S., He, H.: Dual encoding u-net for retinal vessel segmentation.
  In: International Conference on Medical Image Computing and Computer-Assisted
  Intervention. pp. 84--92. Springer (2019)

\bibitem{wang2020rvseg}
Wang, W., Zhong, J., Wu, H., Wen, Z., Qin, J.: Rvseg-net: An efficient feature
  pyramid cascade network for retinal vessel segmentation. In: International
  Conference on Medical Image Computing and Computer-Assisted Intervention. pp.
  796--805. Springer (2020)

\bibitem{xu2020boosting}
Xu, R., Liu, T., Ye, X., Lin, L., Chen, Y.W.: Boosting connectivity in retinal
  vessel segmentation via a recursive semantics-guided network. In:
  International Conference on Medical Image Computing and Computer-Assisted
  Intervention. pp. 786--795. Springer (2020)

\bibitem{xue2020cascaded}
Xue, C., Deng, Q., Li, X., Dou, Q., Heng, P.A.: Cascaded robust learning at
  imperfect labels for chest x-ray segmentation. In: International Conference
  on Medical Image Computing and Computer-Assisted Intervention. pp. 579--588.
  Springer (2020)

\bibitem{you2011segmentation}
You, X., Peng, Q., Yuan, Y., Cheung, Y.m., Lei, J.: Segmentation of retinal
  blood vessels using the radial projection and semi-supervised approach.
  Pattern recognition  \textbf{44}(10-11),  2314--2324 (2011)

\bibitem{zhang2020befd}
Zhang, M., Yu, F., Zhao, J., Zhang, L., Li, Q.: Befd: Boundary enhancement and
  feature denoising for vessel segmentation. In: International Conference on
  Medical Image Computing and Computer-Assisted Intervention. pp. 775--785.
  Springer (2020)

\bibitem{zhang2020retinal}
Zhang, S., Fu, H., Xu, Y., Liu, Y., Tan, M.: Retinal image segmentation with a
  structure-texture demixing network. In: International Conference on Medical
  Image Computing and Computer-Assisted Intervention. pp. 765--774. Springer
  (2020)

\bibitem{zhang1984fast}
Zhang, T., Suen, C.Y.: A fast parallel algorithm for thinning digital patterns.
  Communications of the ACM  \textbf{27}(3),  236--239 (1984)

\bibitem{zhang2018generalized}
Zhang, Z., Sabuncu, M.R.: Generalized cross entropy loss for training deep
  neural networks with noisy labels. arXiv preprint arXiv:1805.07836  (2018)

\bibitem{zhao2015retinal}
Zhao, Y., Liu, Y., Wu, X., Harding, S.P., Zheng, Y.: Retinal vessel
  segmentation: An efficient graph cut approach with retinex and local phase.
  PloS one  \textbf{10}(4),  e0122332 (2015)

\bibitem{zhuang2018laddernet}
Zhuang, J.: Laddernet: Multi-path networks based on u-net for medical image
  segmentation. arXiv preprint arXiv:1810.07810  (2018)

\end{thebibliography}

\end{document}